\documentclass[aip,apl,a4,reprint,twoside,floatfix,superscriptaddress]{revtex4-1} %as an APL paper
\usepackage[dvipdfm]{graphicx}

\usepackage{bm}
\def\vb#1{{\bm#1}}
\def\v#1{\mathbf{#1}}			%'¾Žš'̃xƒNƒgƒ‹
\def\r{\v{r}} 					% def. of vector "r"
\def\k{\v{k}} 					% def. of vector "k"
\def\del{\partial}

\begin{document}

%\preprint{}

\title{Spin current generation due to mechanical rotation in the presence of  impurity scattering} %Title of paper



\author{Mamoru Matsuo$^{1,2}$,
Jun'ichi Ieda$^{2,3}$, Eiji Saitoh$^{2,3,4}$,
and Sadamichi Maekawa$^{2,3}$ }
\affiliation{%
$^{1}$Yukawa Institute for Theoretical Physics,  Kyoto University, Kyoto 606-8502, Japan \\
$^{2}$The Advanced Science Research Center, Japan Atomic Energy Agency, Tokai 319-1195, Japan \\
$^{3}$CREST, Japan Science and Technology Agency, Sanbancho, Tokyo 102-0075, Japan\\
$^{4}$Institute for Materials Research, Tohoku University, Sendai 980-8577, Japan}

\date{\today}

\begin{abstract}
We theoretically investigate the generation of spin current from a uniformly rotating body with impurity scattering
on the basis of the spin dependent transport equation. 
The spin current is created mainly in the radial direction when a weak magnetic field, 
where the cyclotron frequency is smaller than the inverse of the relaxation time due to the impurity scattering,
 is applied parallel to the rotation axis. 
Spin accumulation is estimated by solving the spin diffusion equation.
We show that the inverse spin Hall effect can be used to detect the spin current induced by rotation.
\end{abstract}


\maketitle %\maketitle must follow title, authors, abstract and \pacs
In 1915, Albert Einstein, Wander Johannes de Haas and Samuel Jackson Barnett discovered 
the coupling of magnetism and rotational motion.\cite{Einstein-deHaas1915, Barnett1915}  
They measured the gyromagnetic ratio and the anomalous $g$ factor of electrons 
before the dawn of the modern quantum physics. 
Recent development of  nano-processing technologies has led 
to detect the effects of mechanical rotation on nanostructured magnetic systems.\cite{Wallis2006,Zolfagharkhani2008}
Theoretical studies on the coupling of magnetism and rotation have been performed 
in consideration of nanoscale systems.\cite{Mohanty2004,Kovalev2007,Bretzel2009,Jaafar2009}

Spin-dependent transport phenomena in magnetic nanostructures are of great interest in the field of spintronics, 
which involves the study of ``spin current'', a flow of spins.\cite{MaekawaEd2006} 
The coupling of the magnetization and spin current is one of the hottest areas in spintronics such as 
spin transfer torque,\cite{Slonczewski1996,Berger1996} spin pumping,\cite{Tserkovnyak2002} 
and spin motive force.\cite{Barnes2007}
Recently, the authors constructed the fundamental Hamiltonian with the direct coupling of spin current and mechanical rotation
from the general relativistic Dirac equation, 
and predicted the generation of spin current from a uniformly rotating in a ballistic system.\cite{Mamoru2011}

In this Letter, we extend the results in the ballistic regime to diffusive one introducing effects of impurity scattering on the generation of spin current by mechanical rotation. Combining the conventional transport equation of non-equilibrium steady states with the spin dependent semi-classical equation of electrons in a uniformly rotating frame, we show that the spin current is generated in the radial direction in a weak magnetic field. 
The spin accumulation at the edges of a Pt film attached to a uniformly rotating disk is estimated using 
a spin diffusion equation with a spin source term due to the mechanical rotation. 
It is shown that the inverse spin Hall voltage can be measured 
as a signal of spin current induced by mechanical rotation
in the larger sample compared to the spin diffusion length. 

%%%%%%%

Spin dependent transport in a system with strong spin-orbit interaction can be described as a set of the semi-classical equations\cite{Xiao2010}
\begin{eqnarray}
\dot{\v{r}} = \v{v} + \v{v}_{\vb{\sigma}}, \, \hbar \dot{\v{k}} = -e \left( \v{E} +  \dot{\v{r}} \times \v{B} \right) \label{SemiClassical}
\end{eqnarray}
where $\v{r}$ is the position vector of an electron, $\v{v}=\hbar\k/m$ the normal velocity, $\v{v}_{\vb{\sigma}}$ the anomalous velocity, $\hbar$ Planck constant, $e$ the electron charge, $m$ the electron mass, $\v{k}$ the lattice momentum, $\v{E}$ and $\v{B}$ are electric and magnetic fields. The anomalous velocity originating from the spin-orbit interaction derived from the Dirac equation in an inertial frame is written as $\v{v}_{\vb{\sigma}} = (e\lambda/\hbar)\vb{\sigma} \times \v{E}$ with spin-orbit coupling $\lambda=\hbar^{2}/4m^{2}c^{2}$ and the Pauli matrix $\vb{\sigma}$.

In a uniformly rotating frame, the anomalous velocity is given by\cite{Mamoru2011}
\begin{eqnarray}
\v{v}_{\vb{\sigma}} &=&  \frac{e  \lambda  }{\hbar } \vb{\sigma} \times ( \v{E} + (\vb{\Omega} \times \v{r}) \times \v{B})
\end{eqnarray}
with the rotation frequency vector $\vb{\Omega}$.

\begin{figure}[tbp]
\begin{center}
\includegraphics[scale=0.6]{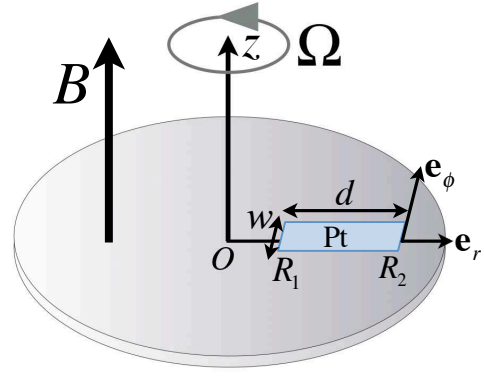}
\end{center}
\caption{A Pt film is attached to a rotating disk. An external magnetic field is applied parallel to a rotation axis.  }
\label{setup}
\end{figure}

We consider the spin current generation in a rotating non-magnetic conductor with a large spin-orbit coupling such as Pt
in the presence of spin-independent impurity scattering. 
Although the scattering does not depend on spins, the electron distribution function depends on spins, owing to the spin-dependence of the semi-classical equations (\ref{SemiClassical}). 
Thus, the transport equation of non-equilibrium steady states is written as
\begin{eqnarray}
\dot{\r} \cdot \frac{\del f_{\vb{\sigma}}}{\del \r} + \dot{\v{k}} \cdot \frac{\del f_{\vb{\sigma}}}{\del \v{k}}  = - \frac{ f_{\vb{\sigma}} - f_{0} }{ \tau }.
\end{eqnarray}
Here, $f_{\vb{\sigma}}=f_{\vb{\sigma}}(\v{r},\v{k})$ is electron distribution function, $f_{0}=f_{0}(\varepsilon)$ the Fermi-Dirac distribution function, and $\tau$ the relaxation time due to the impurity scattering.
In the case of $\v{E}=0$,
the solution of the transport equation is
\begin{eqnarray}
&&f_{\vb{\sigma}} = f_{0} + e\v{v} \cdot   \tau \frac{ \v{E}_{\sigma} + \tau \vb{\omega}_{c} \times \v{E}_{\sigma} }{1+ ( \tau \omega_{c} )^{2}  } \frac{ \del f_{0} }{ \del \varepsilon  }, \\ 
&&\v{E}_{\sigma}=\v{v}_{\vb{\sigma}} \times \v{B} 
= \frac{e\lambda}{\hbar} [\vb{\sigma} \times ((\vb{\Omega} \times \r ) \times \v{B})]\times \v{B}
\end{eqnarray}
where $\vb{\omega}_{c} = e \v{B}/m$ is the cyclotron freaqueancy.
When $\v{B}=(0,0,B)$ and $\vb{\Omega}=(0,0,\Omega)$,  
the spin current induced by the mechanical rotation with impurity scattering can be estimated to be 
$\v{J}_{s}= -e {\rm Tr} [ \sigma_{z} \int d\v{k} f_{\vb{\sigma}}(\r,\k) \v{\dot{r}}]$. Thus, we obtain the explicit form of the spin current in a rotating disk:
\begin{eqnarray}
&& \v{J}_{s}(R) = J_{s}^{r}(R) \v{e}_{r} + J_{s}^{\phi}(R) \v{e}_{\phi}, \\
&& J_{s}^{r}   =  \frac{ \tau \omega_{c} }{1+ (\tau \omega_{c})^{2} } J_{s}^{0}, \,
J_{s}^{\phi}  =  \frac{ (\tau \omega_{c})^{2} }{1+ (\tau \omega_{c})^{2} } J_{s}^{0} \label{Js:r-phi}
\end{eqnarray}
where $\v{e}_{r}(\v{e}_{\phi})$ is the unit vector of the radial(azimuthal) direction as shown in Fig. {\ref{setup}}, and $J_{s}^{0} $ is given by 
\begin{eqnarray}
J_{s}^{0} (R)= 2ne \kappa \omega_{c} R. \label{Js0}
\end{eqnarray}
Here, $n$ is the electron number density, $R$ the distance from the center of the rotation,  and $\kappa= \lambda k_{F}^{2} \hbar \Omega/\varepsilon_{F}$ with Fermi wave vector $k_{F}$ and Fermi energy $\varepsilon_{F}$.

In the limit of large $\tau \omega_{c}$,  Eq. (\ref{Js:r-phi}) reproduces the result in a ballistic regime as $J_{s}^{r}=0$ and $J_{s}^{\phi}=J_{s}^{0}$.\cite{Mamoru2011} 
The formulae (\ref{Js:r-phi}) and (\ref{Js0}) show that the mechanically induced spin current is proportional to 
the spin-orbit coupling $\lambda$ and the rotation frequency $\Omega$,
whereas $J_{s}^{r} \propto B^{2}$ and $J_{s}^{\phi} \propto B^{3}$.  
The spin-orbit coupling $\lambda$ is enhanced in metals and semiconductors such as Pt,\cite{Vila2007,Guo2009,Gu2010} where we use $\lambda k_{F}^{2}=0.59.$\cite{Vila2007}
The rotation frequency is chosen to be $\Omega = 1$kHz, 
where the ultra high speed rotor works stably even in the presence of frictions by the contact measurement tools.\cite{Ono2009}
This rotor is suitable for the detection of the generation of spin current by mechanical rotation in Pt as discussed below.
Thus, the dimensionless parameter becomes $\kappa \approx 10^{-13}$.
For Pt at room temperature in an external magnetic field $B =$ 1T, 
we have $\tau \omega_{c} = 10^{-3}$.
In this condition, the radial component of spin current $J_{s}^{r}$ becomes much larger than the azimuthal component $J_{s}^{\phi}$ because of Eq. (\ref{Js:r-phi}). In the case of $R=100\mbox{mm}$, 
we obtain $J_{s}^{r}\approx 10^{5} \mbox{A/m}^{2}$ and $J_{s}^{\phi}\approx 10^{2} \mbox{A/m}^{2}$.

\begin{figure}[tbp]
\begin{center}
\includegraphics[scale=0.5]{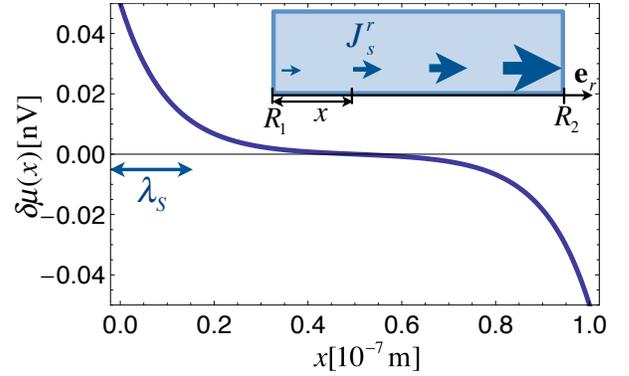}
\end{center}
\caption{In a weak magnetic field, $\omega_{c}\tau \ll 1$, the spin current due to the mechanical rotation is generated in the radial direction. This spin current is included into the spin diffusion equation as a spin source term.
The spin accumulation $\delta \mu$ along the radial direction in Pt is plotted as a function of the distance from the edge, $x$. The sample size $d=100$nm, the distance between the sample edge and the rotation axis $R_{1}=50$mm.  }
\label{SA}
\end{figure}

We consider the spin accumulation in a nanoscale Pt film which is rotated by a high speed rotator in a weak magnetic regime, $\omega_{c}\tau \ll 1$.
The Pt film with width $w$ and length $d$ is attached to a disk and the distance between the edges and the rotation axis are $R_{1}$ and $R_{2}$ as shown in  Fig. \ref{setup} .  
In the film, the radial spin current is induced when the disk is rotating. 
In the weak magnetic regime, the spin current mainly flows in the radial direction as mentioned above. 
When keeping the rotation frequency constant, the non-equilibrium steady state is achieved. 
In this situation,  
the spin accumulation near the edges can be estimated by the spin diffusion equation with the source term originating from mechanical rotation as follows:
\begin{eqnarray}
\nabla^{2} \delta \mu = \frac{1}{\lambda_{s}^{2}} \delta \mu - e  \rho\,  \mbox{div} J_{s} \label{SpinDiff}
\end{eqnarray}
where $\delta \mu, \lambda_{s}$, and $ \rho$ are spin accumulation, spin diffusion length, and resistivity. 
The total spin current in the film $J_{s}^{\rm tot}$ consists of the diffusive spin current $(1/2e\rho)\nabla \delta \mu$ and the mechanically induced current $J_{s}^{r,\phi}$.
The analytic solution of Eq. (\ref{SpinDiff}) in the cylindrical polar coordinates with the boundary conditions, $J_{s}^{\rm tot}=0$ at the edges, is given by the superposition of the modified Bessel function of the first and second kind. 
Choosing $d=100$nm, $R_{1}=50$mm, $\Omega=1$kHz, $B=1$T, $\lambda_{s}=14$nm, and $\rho=12.8 \times 10^{-8}\Omega\cdot$m for Pt,\cite{Steenwyk1997}  we obtain the spin accumulation near the edges $\delta \mu(R_{1}) \approx \delta \mu(R_{2}) \approx 0.05$nV as shown in Fig. \ref{SA}.

%%%%%%%

\begin{figure}[tbp]
\begin{center}
\includegraphics[scale=0.5]{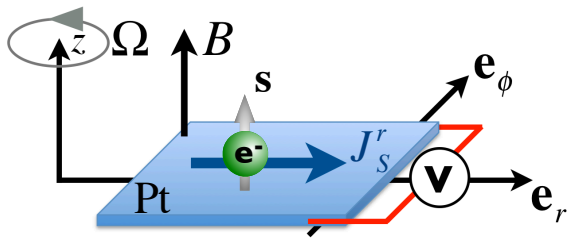}
\end{center}
\caption{The radial spin current is converted to the inverse spin Hall voltage in the azimuthal direction $\v{e}_{\phi}$. }
\label{ISHE}
\end{figure}

Let us consider a detection method of the spin current due to mechanical rotation 
 by means of the inverse spin Hall effect.\cite{Saitoh2006}
 The inverse spin Hall effect converts a spin current into an electric voltage via the spin-orbit interaction 
as $\v{E}_{{\rm ISHE}} \propto \v{J}_{s} \times \v{s}$, where $\v{E}_{{\rm ISHE}}, \v{J}_{s}$ and $\vb{s}$ are the electric field due to the inverse spin Hall effect, the spatial direction of the spin current, and the spin-polarization vector of the spin current, respectively.
In the present Pt sample attached to a rotating disk, the voltage due to the inverse spin Hall effect is generated in the azimuthal direction.
The voltage by a voltmeter, attached to both sides of the film,
gives us the evidence of the generation of radial spin current due to the mechanical rotation (see Fig. \ref{ISHE}).
In the Pt film, the $z$-polarized spin current flowing in the radial direction converts the inverse spin Hall voltage in the azimuthal direction.

We note that the voltage signal induced by the inverse spin Hall effect is completely different from 
the electromotive force produced by the acceleration known as Tolman-Stewart effect.\cite{Tolman1916}
The voltage originating from the Tolman-Stewart effect is created in the radial direction because of
the acceleration of the rotational motion,
whereas the inverse spin Hall voltage is induced in the azimuthal direction as indicated above.

The inverse spin Hall voltage $V_{{\rm ISHE}}$ is estimated by $V_{{\rm ISHE}} = \Theta w \rho \langle  J_{s}^{r} \rangle $ with the spin Hall angle $\Theta$, sample width $w$, resistivity $\rho$ and the average of the radial spin current $ \langle  J_{s}^{r} \rangle $ which is calculated by
\begin{eqnarray}
\langle  J_{s}^{r} \rangle = \frac{1}{d} \int_{R_{1}}^{R_{2}} J_{s}^{r}dR =\frac{ \omega_{c}^{2}\tau ne \kappa  (R_{1}+R_{2})}{1+ (\omega_{c}\tau )^{2}}. \label{AvJsr}
\end{eqnarray}
From Eq. (\ref{AvJsr}), we have the voltage $V_{\rm ISHE} \propto B^{2} \Omega$, 
which shows the sign of the voltage changes only when the rotation axis is reversed.
In the case of $w=d=10$mm, $\omega_{c} \tau=0.01$, $\Theta=0.01$ and $\rho=12.8\times 10^{-8} \Omega\cdot$m, 
we obtain the average of the radial spin current $\langle  J_{s}^{r} \rangle \approx 10^{5} {\rm A/m^{2} }$ and
the voltage $V_{{\rm ISHE}} \approx 1 \mu \mbox{V}$.

%%%%conclusion
In conclusion, we have investigated the generation of spin current due to mechanical rotation
in the presence of impurity scattering.
The explicit form of spin current is obtained by combining a conventional transport equation of non-equilibrium steady states with the semi-classical electron equation with the spin-dependent velocity in a uniformly rotating frame.
It is shown that the spin current is mainly generated in the radial direction in a weak magnetic regime. 
The spin accumulation in a Pt film attached to the uniformly rotating disk with an external magnetic field is calculated
by solving the spin diffusion equation along the radial direction with the spin source term.
We have proposed that the inverse spin Hall voltage can be used as a signal of the generation of spin current from the rotating body.

% If in two-column mode, this environment will change to single-column format so that long equations can be displayed. 
%\begin{widetext}
%$$\mbox{put long equation here}$$
%\end{widetext}

% Figures should be put into the text as floats. 
% Use the graphics or graphicx packages (distributed with LaTeX2e).
% Put the label that you will use with \ref{} command in the braces of the \label{} command.
%
% \begin{figure}
% \end{figure}

% \begin{table}
% \end{tabular}
% \end{table}

\begin{acknowledgments}
The authors are grateful to S. Takahashi, M. Ono, T. Ono, and Z. Fisk for valuable discussions.  
This work was supported by a Grant-in-Aid for Scientific Research from MEXT, Japan  
and the Next Generation Supercomputer Project, Nanoscience Program from MEXT, Japan.
\end{acknowledgments}

\end{document}